\documentclass[showpacs,aps,pra,superscriptaddress,twocolumn,amsmath,amssymb]{revtex4}

\usepackage{graphicx}
\usepackage{dcolumn}
\usepackage{bm}
\usepackage{amsmath,epsfig}
\usepackage{tabularx}

\newcommand{\abs}[1]{\left\vert#1\right\vert}
\newcommand{\ket}[1]{\left\vert#1\right\rangle}
\newcommand{\bra}[1]{\left\langle#1\right\vert}
\newcommand{\braket}[2]{\left\langle#1\vert#2\right\rangle}

\begin{document}

\author{Michele Tumminello}
\affiliation{Dipartimento di Fisica e Tecnologie Relative,
Universit\`a di Palermo, Viale delle Scienze, I-90128 Palermo, Italy}

\email {tumminello@lagash.dft.unipa.it}

\author{Francesco Ciccarello}
\affiliation{Dipartimento di Fisica e Tecnologie Relative,
Universit\`a di Palermo, Viale delle Scienze, I-90128 Palermo,
Italy} \affiliation{Consorzio Nazionale Interuniversitario per le Scienze Fisiche della Materia , Italy} \affiliation{NEST and Dipartimento di Scienze Fisiche ed
Astronomiche, Universit\`{a} di Palermo, Via Archirafi 36, I-90123
Palermo, Italy}

\begin{abstract}
We present a scheme for conditionally teleporting an unknown atomic
state in cavity QED which requires two atoms and one cavity mode.
The translational degrees of freedom of the atoms are taken into
account using the optical Stern-Gerlach model. We show that
successful teleportation with probability 1/2 can be achieved
through local measurements of the cavity photon number and atomic
positions. Neither direct projection onto highly entangled
states nor holonomous interaction-time constraints are required.
\end{abstract}


\title{Teleportation of atomic states via position measurements}

\maketitle

\section{Introduction}

Quantum entanglement, maybe the most intriguing feature of quantum
mechanics \cite{epr}, is a powerful resource for quantum information
processing tasks \cite{nielsen}.

An outstanding application of entanglement is the teleportation of
an unknown qubit, the unit of quantum information, between two
systems. In the seminal paper by Bennett \emph{et al.}
\cite{bennett}, a quantum state is transferred from qubit $A$ to
qubit $B$ using an \emph{ancilla}, e.g. a third auxiliary qubit $C$.
Qubits $B$ and $C$ are initially prepared in an entangled state. A
Bell measurement on $A$ and $C$ is then made. Depending on the
outcome of such measurement, a suitable unitary transformation on
$B$ is performed in order to reconstruct the initial quantum state
of $A$. Teleportation is successful with probability $1$. Soon
after the proposal by Bennett \emph{et al.}, quantum teleportation
has received considerable attention culminated in its experimental
demonstration in a number of works
\cite{bowm-boschi,NMR,natphys,nature}.

Cavity QED systems -- where Rydberg atoms couple to the quantized
electromagnetic (e.m.) field of a superconductive cavity
\cite{raimond} -- have received considerable attention during the
last years \cite{teleportation-QED}. Cavity QED systems have been
proposed for implementing teleportation protocols of internal
quantum states between atoms, a task which is particularly
attractive especially after its experimental proof for trapped ion
systems \cite{teleportation-ions}. Generally speaking, in such
cavity QED schemes a quantum internal state is teleported between
two atoms via coherent interaction with cavity field modes and/or
auxiliary atoms which act as quantum channels.

Quite recently, efforts have been done for achieving teleportation
without direct projections onto Bell states
\cite{vaidman,de-almeida,zheng,ye-guo,cardoso}. In particular, Zheng
has proposed a scheme for approximately teleporting an unknown
internal state between two atoms which successively interact with a
cavity mode according to the Jaynes-Cummings Hamiltonian
\cite{zheng}. The probability of success is 1/4 and only
measurements of product states are required. Ye and Guo have
presented another scheme that does not require projections onto Bell
states and makes use of three atoms and a single-mode cavity field
out of resonance \cite{ye-guo}. The atom-atom coupling via the
virtual excitations of the cavity field is exploited for teleporting
a quantum state between two atoms with probability of success 1/2.
Ye and Guo presented their work in terms of a ``no Bell-state
measurement scheme". This parlance was later criticized in a comment
by Chhajlany and Wójcik \cite{comment-ye-guo} who showed how the
scheme by Ye and Guo, despite its use of local measurements, in fact
relies on Bell state measurements. Protocols of this sort are indeed
more properly classified as methods to achieve teleportation without
requiring \emph{direct} projections onto Bell states
\cite{ye-guo-reply}. Noticeably, both the schemes by Zheng
\cite{zheng} and Ye and Guo \cite{ye-guo} require precise tuning of
the atom-cavity field interaction time.

To our knowledge, no cavity QED-teleportation scheme has so far
accounted for the translational dynamics of atoms flying through a
cavity. Indeed, the spatial structure of the quantum e.m. field
along the $x$-cavity axis affects the internal dynamics of a flying
atom. This leads to an atom-field coupling constant which in fact
depends on the atomic translational degrees of freedom along the
$x$-direction. Such circumstance -- taking place whenever the atomic
wavepacket has a width non negligible with respect to the field
wavelength -- has been shown to give rise to a number of observable
phenomena such as optical Stern-Gerlach effect \cite{SGE},
self-induced transparency \cite{schlicher}, modulation of the atomic
decay in a damped cavity \cite{wilkens}, non-dissipative damping of
the Rabi oscillations \cite{Vag-Cus,which-path}.

It is clear that the involvement of the translational degrees of
freedom introduces non-dissipative decoherence in the atom-field
dynamics. Such effect, stemming from the entanglement between the
atom-field system and the atomic translational degrees of freedom, has been
shown to spoil the non-local correlations between two atoms which
successively interact with the same cavity mode
\cite{epl-2atoms,epjd}. Accordingly, the inclusion of the
translational dynamics is thus expected to decrease the efficiency
of those teleportation protocols relying on the coherent atom-cavity
mode coupling.

However, a different perspective can be adopted. Indeed, one may
wonder whether such additional degrees of freedom could be
fruitfully exploited as a resource for attaining efficient atomic
teleportation provided that measurements of the atomic positions are
performed. According to such a scenario, the atomic translational
degrees of freedom play the role of further quantum channels able to
transfer information between the internal degrees of freedom of
different atoms.

A crucial motivation in the search for such a teleportation protocol
is that, according to the optical Stern-Gerlach model, the
wavefunction of a two-level atom entering a cavity generally splits
into a set of deflected wavepackets, each corresponding to a
different atom-field dressed state \cite{Vag-Cus,vaglica95}.
For an increasing atom-cavity interaction time, such
outgoing wavepackets become more and more distinguishable up to the
point that \emph{which-path} information becomes accessible
\cite{which-path}. This information is used in our protocol
for attaining conditional transfer of quantum information between
two atoms which successively interact with the same cavity mode.
This is indeed the central mechanism underlying the physics
presented in this work.

In this paper, we consider two atoms which successively enter the
same cavity in either a nodal or antinodal region of the
corresponding field mode. Each atom interacts with such mode
according to the optical Stern-Gerlach Hamiltonian. This can be
approximated as a linear (quadratic) expansion in the atomic
position along the cavity axis when a nodal (antinodal) region is
considered. Both the atoms are assumed to enter the cavity in a
given minimum uncertainty Gaussian wave packet with the target atom
and the resonant mode initially in the excited and vacuum state,
respectively. We show that conditional teleportation of an internal
atomic state can be achieved by local measurements of the atomic
positions, the cavity photon-number and the internal state of the
atom whose state is to be transmitted. No direct Bell-state
measurement is required. We thus prevent the projection of our
two-atoms system onto highly entangled subspaces, therefore avoiding
the need of (in general quite difficult) joint measurements. This is
a major advantage of teleportation schemes that do not rely on
direct Bell-state measurements. Furthermore, at variance with other
cavity-QED protocols which work without direct Bell-state
measurements \cite{zheng,ye-guo}, no holonomous constraints on the
atom-cavity interaction times are required. It only suffices that
the time of flight of each atom inside the cavity is long enough in
order for the outgoing deflected wavepackets to be distinguished
with reasonable approximation.
We show that successful teleportation of an atomic
state can be attained with probability $1/2$.

This paper is organized as follows. In Sec. \ref{syst-appr}, we
introduce the system and the Hamiltonian both in the nodal and in
the antinodal case. In Sec. \ref{QT-section}, the main part of this
work, we describe the teleportation scheme. A relevant property the
protocol relies on is the \emph{which-path} information about the
outgoing atomic wave packets. The conditions allowing this
information to be accessible are reviewed and discussed in Sec.
\ref{orthogonality}. Finally, in Sec. \ref{conclusions}, we draw our
conclusions.

\section{System and Approach} \label{syst-appr}

We consider two identical two-level atoms, labeled 1 and 2, of mass
$m$ and Bohr frequency $\omega$. The atoms interact in succession
with the e.m. field of the same e.m. cavity. We assume that the
velocity of each atom along the $z$-direction (orthogonal to the
$x$-cavity axis) is large enough that the motion along the $z$-axis
is not affected by the cavity field and can be treated
classically. Denoting by $a$ and $a^{\dag}$ the annihilation and
creation operators of the cavity field and assuming the resonance
condition, the free Hamiltonian $H_0$ can be written as
\begin{equation}\label{H0}
H_0=\sum_{i=1,2}\left[\frac{\hat{p}_i^2}{2m}+\hbar \omega
S_{z,i}\right]+\hbar \omega a^{\dag}a \,,
\end{equation}
where -- for each atom $i=1,2$ -- $S_{z,i},S_{\pm,i}$ are the usual
spin-1/2 operators and $\hat{p}_i=-i\hbar(d/dx_{i})$ is the
$x$-component of the momentum operator. In the Rotating Wave
Approximation, each atom $i$ couples to the cavity field according
to the interaction Hamiltonian
\begin{equation}\label{Hif}
H_{if}=\hbar \varepsilon \sin (k\hat{x}_{i})
\left(a^{\dag}S_{-,i}+aS_{+,i}\right) \,\,\,\,(i=1,2)
\end{equation}
with $k$ and $\varepsilon$ standing for the wave number of the e.m.
mode and the atom-field coupling constant, respectively, and where
$\hat{x}_{i}$ is the $i$th atomic position operator along the cavity
axis.

Hamiltonian (\ref{Hif}) accounts for the spatial structure of the
e.m. field along the $x$-cavity axis. Rigorously speaking, it should
be mentioned that the atom-field coupling constant has also a
spatial structure along both the $y$ and $z$-axes perpendicular to
the cavity axis. Such structure, having a gaussian profile of the
form $\exp [-(y^2+z^2)/w_0^2]$ ($w_0$ cavity waist)
\cite{carmichael}, is neglected by the optical Stern-Gerlach
interaction Hamiltonian (\ref{Hif}). Concerning the $z$-axis, the
large atomic velocity along such direction indeed ensures that each
flying atom is insensitive to the cavity field and thus to its
structure along such axis. On the other hand, we assume to be in the
regime such that $w_0 \gg 2\pi/k$. In this case, it is enough to
take into account only the $x$-structure of the e.m. field, assuming
a uniform spatial dependence on the transversal direction. Such a
regime is a feasible one given that microwave cavities having a
value of $w_0 \gg 2\pi/k$ are quite common (see e.g. \cite{haroche}
where $w_0$ is as large as 6 mm).

When both the atoms enter the cavity in a nodal region of the cavity
mode with the width $\sigma_{x_{i}}$ of their respective wavepackets
small enough compared to $2\pi/k$ ($\sigma_{x_{i}}\ll 2\pi/k$),
$H_{i}$ can be approximated as a linear expansion in the atomic
position
\begin{equation}\label{Hif-nodal}
H_{iN}=\hbar \varepsilon k\, \hat{x}_{i}
\left(a^{\dag}S_{-,i}+aS_{+,i}\right) \,,
\end{equation}
while in an antinodal region it takes the form
\begin{equation}\label{Hif-antinodal}
H_{iA}=\hbar \varepsilon \left(1-\frac{k^2\hat{x}_{i}^2}{2}\right)
\left(a^{\dag}S_{-,i}+aS_{+,i}\right) \,.
\end{equation}
In Eqs. (\ref{Hif-nodal}) and (\ref{Hif-antinodal}),
$\hat{x}_{i}$ stands for the atomic position operator of the $i$th atom with
respect to a nodal point and an antinodal point, respectively.

At time $t=0$, atom 1 enters the cavity and interacts with the field
for a time $t_{1}$. At a later time $t_{2}>t_{1}$, atom 2 enters the
cavity and couples to the field state modified by the first atom. At
time $t_{3}>t_{2}$ atom 2 exits the cavity. At times $t\geq t_{3}$
both the atoms are therefore out of the cavity and evolve freely. In
the interaction picture, the Hamiltonian at all times in a nodal
region of the cavity field, reads
\begin{eqnarray}\label{H-nodal}
  H_{N}^{I}(t)&=&\hbar\varepsilon k\left(\hat{x}_{1}+\frac{\hat{p}_{1}}
  {m}t\right)\mu_{t}(0,t_{1})u_{1} \nonumber \\&+& \hbar\varepsilon k\left(\hat{x}_{2}+\frac{\hat{p}_{2}}{m}t\right)
  \mu_{t}(t_{2},t_{3})u_{2} \,,
\end{eqnarray}
where we have introduced the atom-field operators
$u_{i}=a^{\dag}S_{-,i}+aS_{+,i}$ and where the time interval during
which each atom interacts with the cavity mode is accounted for
through the function $\mu_{t}(t',t'')=\theta(t-t')-\theta(t-t'')$,
$\theta(t)$ being the usual Heaviside function.

In an antinodal region of the cavity field, the Hamiltonian in the
interaction picture takes the form
\begin{eqnarray}\label{H-antinodal}
H_{A}^{I}(t)&=&\hbar\varepsilon
\left[1-\frac{k^2}{2}\left(\hat{x}_{1}+\frac{\hat{p}_{1}}
  {m}t\right)^2\right] \mu_{t}(0,t_{1})u_{1}\nonumber \\&+& \hbar\varepsilon \left[1-\frac{k^2}{2}\left(\hat{x}_{2}+\frac{\hat{p}_{2}}
  {m}t\right)^2\right]
  \mu_{t}(t_{2},t_{3})u_{2}\,.
\end{eqnarray}
Of course, in the time interval $[t_1,t_2]$ and for $t\geq t_3$ both
$H_{N}^{I}(t)$ and $H_{A}^{I}(t)$ vanish since no atom is inside the
cavity. The Hamiltonian operators of Eqs. (\ref{H-nodal}) and
(\ref{H-antinodal}) can be used to derive the exact dynamics of a
given initial state of the two-atom-field system at times $t\geq
t_3$. This is accomplished through the respective evolution
operators $U_{\alpha}^{I}(t\geq t_{3})$
\begin{equation}\label{Ualpha}
U_{\alpha}^{I}(t\geq
t_{3})=T\,\exp\left[-\frac{i}{\hbar}\int_{0}^{t_3}
H_{\alpha}^{I}(t)dt\right]\,\,\,\,\,\,\,(\alpha=N,A)
\end{equation}
with $T$ standing for the time-ordering operator and where the second integration bound is due to the fact that
$H_{\alpha}^{I}=0$ for $t\geq t_{3}$.

Due to the fact that atom 2 enters the cavity after atom 1 has come
out of it, it is possible to split up $U_{\alpha}^{I}(t\geq t_{3})$
into the product of two evolution operators $U_{\alpha,1}^{I}(t\geq
t_{3})$ and $U_{\alpha,2}^{I}(t\geq t_{3})$ ($\alpha=N,A$). Each
operator $U_{\alpha,i}^{I}(t\geq t_{3})$ only affects the dynamics
of atom $i$. In formulae (from now on, whenever unnecessary, the
time argument ``$(t\geq t_{3})$" and/or the apex ``$I$"  in the
evolution operators will be omitted)
\begin{equation}\label{U-alpha}
U_{\alpha}=U_{\alpha,2} \cdot
U_{\alpha,1}\,\,\,\,\,\,\,\,\,\,\,\,(\alpha=N,A)
\end{equation}
with
\begin{eqnarray}
U_{\alpha,1}=T\,\exp\left[-\frac{i}{\hbar}\int_{0}^{t_1} H_{\alpha}^{I}(t)dt\right]=U_{\alpha,1}(\hat{x}_1,\hat{p}_1,u_1), \label{Ualpha1}\\
U_{\alpha,2}=T\,\exp\left[-\frac{i}{\hbar}\int_{t_2}^{t_3}
H_{\alpha}^{I}(t)dt\right]=U_{\alpha,2}(\hat{x}_2,\hat{p}_2,u_2),\label{Ualpha2}
\end{eqnarray}
where in the right-hand side of both equations we have explicitly indicated the operators each $U_{\alpha,i}$ depends on according to
Eqs. (\ref{H-nodal}) and (\ref{H-antinodal}).

\section{Teleportation scheme} \label{QT-section}

We denote the ground and excited states of the $i$th atom by
$\ket{g_{i}}$ and $\ket{e_{i}}$, respectively. Assume that atom 2 is
the one whose initial internal state, say $\ket{\alpha_{2}}$, is to
be teleported. Such state is written as
\begin{equation}\label{state-teleported}
\ket{\alpha_{2}}=\cos
\frac{\vartheta}{2}\ket{e_2}+e^{i\varphi}\sin\frac{\vartheta}{2}\ket{g_2}
\end{equation}
with $\vartheta\in[0,\pi]$ and $\varphi\in[0,\pi]$.

By indicating the Fock states of the cavity field as $\ket{n}$
($n=0,1,...$), we consider the following initial state of the
system:
\begin{equation}\label{initial-state}
\ket{\Psi(0)}=\ket{\varphi_1(0)}\ket{e_1}\,\, \ket{\varphi_2(0)}\ket{\alpha_{2}}\,\,\ket{0}\,,
\end{equation}
where $\ket{\varphi_i(0)}$ (associated with each atom $i=1,2$) is a Gaussian
wavepacket of minimum uncertainty, such that the product between the
initial position and momentum widths fulfills
$\sigma_{x_{i}}\cdot\sigma_{ p_{i}}= \hbar/2$.

Consider now the usual dressed states of the $i$th atom
$\ket{\chi_{n,i}^{\pm}}=\left(\ket{e_{i}}\ket{n}\pm
\ket{g_{i}}\ket{n+1}\right)/\sqrt{2}$ ($n=0,1,...$). These states
are eigenstates of the $u_i$ operators since
$u_i\ket{\chi_{n,i}^{\pm}}=\pm \sqrt{n+1} \ket{\chi_{n,i}^{\pm}}$
(while $u_{i}\ket{g_{i}}\ket{0}=0$). The dressed states together
with $\ket{g_{i}}\ket{0}$ ($i=1,2$) represent an orthonormal basis
of the corresponding Hilbert space. It is important to notice that
$u_i$ commutes with $U_{\alpha,i}$ according to Eqs. (\ref{Ualpha1}
and \ref{Ualpha2}) and the corresponding Hamiltonian operators of
Eqs. (\ref{H-nodal} and \ref{H-antinodal}). It follows that the
effective representation $U_{\alpha,i}^{(n,\pm)}$ of $U_{\alpha,i}$,
as applied to a dressed state $\ket{\chi_{n,i}^{\pm}}$, is obtained
by simply replacing $u_i$ with $\pm \sqrt{n+1}$ in Eqs.
(\ref{Ualpha1}) and (\ref{Ualpha2}). This yields
\begin{equation}\label{U-Ni-eff}
U_{\alpha,i}^{(n,\pm)}=U_{\alpha,i}(\hat{x}_i,\hat{p}_i,\pm
\sqrt{n+1})\,\,\,\,\,\,\,(n=0,1,...),
\end{equation}
while the effective representation of $U_{N,i}$ -- as applied to
state $\ket{g_{i}}\ket{0}$ -- reduces to the identity operator for
both the atoms $i=1,2$.

The operators in Eq.~(\ref{U-Ni-eff}) clearly affect only the atomic
translational dynamics and therefore allow to define a family of
atomic translational wavepackets $\ket{\Phi_{\alpha,n,i}^{\pm}}$
according to
\begin{equation}\label{stati_phi}
\ket{\Phi_{\alpha,n,i}^{\pm}}=U_{\alpha,i}^{(n,\pm)}\ket{\varphi_i(0)},
\end{equation}
such that
\begin{equation} \label{stati_phi2}
U_{\alpha,i} \ket{\varphi_i(0)}\ket{\chi_{n,i}^{\pm}}=\ket{
\Phi_{\alpha,n,i}^{\pm}}\ket{\chi_{n,i}^{\pm}}.
\end{equation}


Once the time evolution operator (\ref{U-alpha}) is applied to
$\ket{\Psi(0)}$, the state of the whole system at a time $t \ge
t_{3}$ -- when both the atoms are out of the cavity -- can be
written in the form (from now on, the index $\alpha$ in the $\Phi$ states will
be omitted)
\begin{eqnarray}\label{expansion-t3}
\ket{\psi(t_3)}&=&\ket{\lambda_{0,1}}\ket{\varphi_2(0)}\ket{g_2}\ket{0}\nonumber
\\& &+\sum_{n=0,1}\sum_{\eta=-,+}\left(\ket{\lambda_{n,1}^{\eta}}\ket{\Phi_{n,2}^{\eta}}\ket{\chi_{n,2}^{\eta}}\right),
\end{eqnarray}
where the $\lambda$ states of atom 1 are defined according to
\begin{eqnarray}\label{lambda}
\ket{\lambda_{0,1}}&=&\left(\frac{\ket{\Phi_{0,1}^{+}}+\ket{\Phi_{0,1}^{-}}}{2}\right)\,e^{i\varphi} \sin\frac{\vartheta}{2}\ket{e_1},\\
\ket{\lambda_{0,1}^{\pm}}&=&\left(\frac{\ket{\Phi_{0,1}^{+}}+\ket{\Phi_{0,1}^{-}}}{2\sqrt{2}}\right)\,\cos\frac{\vartheta}{2}\ket{e_1}\nonumber \\
& &\pm  \left(\frac{\ket{\Phi_{0,1}^{+}}-\ket{\Phi_{0,1}^{-}}}{2\sqrt{2}}\right) \,e^{i\varphi}\sin\frac{\vartheta}{2}\ket{g_1}, \\
\ket{\lambda_{1,1}^{\pm}}&=&\left(\frac{\ket{\Phi_{0,1}^{+}}-\ket{\Phi_{0,1}^{-}}}{2\sqrt{2}}\right)\cos\frac{\vartheta}{2}\ket{g_1}.
\end{eqnarray}


The procedure for obtaining state $\ket{\psi(t_3)}$ is detailed in
Appendix \ref{AppendixA}. In what follows, we shall indicate the
time spent inside the cavity by atoms 1 and 2 with $\tau_1=t_2-t_1$
and $\tau_2=t_3-t_2$ respectively. The states
$\ket{\Phi_{n,i}^{\pm}}$ appearing in Eq.~(\ref{expansion-t3})
fulfill the following important property both in the nodal and
antinodal case \cite{which-path,epl-2atoms,epjd}
\begin{eqnarray}\label{cond-stati-phi}
\lim_{\tau_i\rightarrow
\infty}\bra{\Phi_{n,i}^{+}}\Phi_{n,i}^{-}\rangle =0. \label{prop2}
\end{eqnarray}
Such property, together with the features of the outgoing
wavepackets $\ket{\Phi_{n,i}^{+}}$, is discussed in Sec.
\ref{orthogonality}.

According to Eq.~(\ref{cond-stati-phi}), wavepackets
$\ket{\Phi_{n,i}^{+}}$ and $\ket{\Phi_{n,i}^{-}}$ exhibit a
negligible overlap for long enough times of flight $\tau_i$. As
shown in Refs. \cite{epl-2atoms,epjd}, times of flight of the order
of a few Rabi oscillations are sufficient in order to get negligible
overlapping \cite{footnote22}.

Such outstanding circumstance makes it possible to distinguish the
elements of the set of translational states
\{$\ket{\Phi_{n,i}^{\pm}}$\} through measurements of the atomic
positions along the $x$-axis \cite{nota_misura}.

It is straightforward to show that Eq.~(\ref{cond-stati-phi})
implies that all the terms appearing in (\ref{expansion-t3}) are
orthogonal provided that $\tau_1$ and $\tau_2$ are sufficiently
large.

Once the dressed states $\ket{\chi_{n,2}^{\pm}}$ appearing in
Eq.~(\ref{expansion-t3}) are rewritten in terms of states
$\ket{g_2}\ket{n}$ and $\ket{e_2}\ket{n}$, one recognizes the
occurrence of cases where measurements of the photon number, of the
internal state of atom 2 and of the positions of the two atoms can
make atom 1 collapse into the initial internal state of atom 2
[Eq.~(\ref{state-teleported})]. Namely a successful teleportation
can take place. For instance, the projection of $\ket{\psi(t_3)}$
onto the the cavity field state $\ket{1}$ gives
\begin{widetext}
\begin{eqnarray}\label{1photon}
\bra{1}\psi(t_3)\rangle&=&\left[\frac{\left(\ket{\Phi_{0,1}^+}+\ket{\Phi_{0,1}^-}\right)\left(\ket{\Phi_{0,2}^+}-\ket{\Phi_{0,2}^-}\right)}{4}\,\cos\frac{\vartheta}{2}\ket{e_1}+\frac{\left(\ket{\Phi_{0,1}^+}-\ket{\Phi_{0,1}^-}\right)\left(\ket{\Phi_{0,2}^+}+\ket{\Phi_{0,2}^-}\right)}{4}\,e^{i\varphi}\sin\frac{\vartheta}{2}\ket{g_1}\right]\ket{g_2}
\nonumber \\
&
&+\left[\frac{\left(\ket{\Phi_{0,1}^+}-\ket{\Phi_{0,1}^-}\right)\left(\ket{\Phi_{1,2}^+}+\ket{\Phi_{1,2}^-}\right)}{4}\,\cos\frac{\vartheta}{2}\ket{g_1}\right]\ket{e_2}.
\end{eqnarray}
\end{widetext}
This outcome occurs with probability $(3+\cos\vartheta)/8$. Assume
now that a further measurement of the internal state of atom 2 is
made. If the outcome of such measurement is $\ket{e_2}$, atom 1 is
projected onto the ground state $\ket{g_1}$ and thus no
teleportation of the initial state of atom 2 has occurred. The
unconditional probability for this event is calculated to be
$(1+\cos\vartheta)/8$.

However, it can be noticed that if atom 2 is found in the ground
state $\ket{g_2}$ a further measurement of the atomic positions with
outcomes $\ket{\Phi_{0,1}^+}\ket{\Phi_{0,2}^+}$ or
$\ket{\Phi_{0,1}^-}\ket{\Phi_{0,2}^-}$ projects atom 1 into the
state $\ket{\alpha_{1}}=\cos
\frac{\vartheta}{2}\ket{e_1}+e^{i\varphi}\sin\frac{\vartheta}{2}\ket{g_1}$.
This means that state $\ket{\alpha_2}$ of
Eq.~(\ref{state-teleported}) has been in fact teleported into atom
1.

On the other hand, when the wavepackets
$\ket{\Phi_{0,1}^+}\ket{\Phi_{0,2}^-}$ or
$\ket{\Phi_{0,1}^-}\ket{\Phi_{0,2}^+}$ are found (after that the
state $\ket{g_2}$ has been measured) atom 1 collapses into the state
\begin{equation}\label{state-teleported-ruotato}
\ket{\alpha_{1}'}=\cos
\frac{\vartheta}{2}\ket{e_1}-e^{i\varphi}\sin\frac{\vartheta}{2}\ket{g_1}\,,
\end{equation}
which can be easily transformed into (\ref{state-teleported})
through a 180 degree rotation around the $z$-axis in order to
faithfully reproduce the initial state of atom 2 and complete the
teleportation. Of course, rigorously speaking, the
measurements of the atomic positions do not formally correspond to
projections onto states $\ket{\Phi_{0,i}^+}$ and
$\ket{\Phi_{0,i}^-}$. However, due to the discussed orthogonality of
$\ket{\Phi_{0,i}^+}$ and $\ket{\Phi_{0,i}^-}$, such translational
states can be associated with different atomic paths $l^{+}_i$ and
$l^{-}_i$.  The measurements of the atomic positions cause indeed
effective projections on such paths.

Note that the above teleportation scheme, conditioned to the outcome
$\ket{g_2}\ket{1}$, is invariant for a change of each $l^{+}_i$ into
$l^{-}_i$ and vice-versa. This implies that for each atom $i=1,2$
the labeling of the two paths is arbitrary. If both the atoms are
found in a path ``$+$" or in a path ``$-$", atom 1 is projected into
state (\ref{state-teleported}). If the paths of the two atoms have
different signs, regardless of which atom is in which path, state
(\ref{state-teleported-ruotato}) is obtained and the teleportation
process can be finalized once a 180 degree rotation on the internal
state of atom 1 is applied.

In a similar way, it turns out that, when the field vacuum state
$\ket{0}$ is found, the outcome $\ket{g_2}$ cannot transfer the
initial state of atom 2 into atom 1, while successful teleportation
is attained when atom 2 is found to be in the excited state
$\ket{e_2}$. As in the case $\ket{g_2}\ket{1}$, when the atoms are
found in the same quantum path (i.e.  $l^+_1$ and $l^+_2$ or $l^-_1$
and $l^-_2$) the first atom is projected into $\ket{\alpha_1}$.
Again, when different quantum paths are found (i.e. $l^+_1$ and
$l^-_2$ or $l^-_1$ and $l^+_2$) teleportation can be finalized after
a 180 degree rotation around the $z$-axis. Due to conservation of
$\sum_{i=1,2}S_{z,i}+ a^{\dag}a$, no teleportation is possible when
the field is found to be in $\ket{2}$.

\begin{figure}[htbp] \label{figura} {\hbox{
{\includegraphics[scale=0.5]{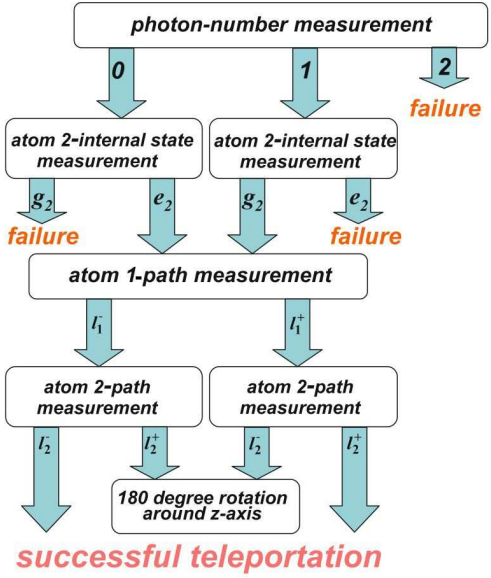}}}}
\caption{\label{schemino}(Color online) Schematic diagram of the
teleportation protocol.}
\end{figure}

All the possible outcomes of the protocol are summarized in
Table \ref{table}. For each case -- corresponding to given outcomes
of the number of photons (1$^{st}$ column), the internal state of
atom 2 (2$^{nd}$ column), and the paths along which the two atoms
are found (3$^{th}$ and 4$^{th}$ columns) -- it is shown
whether or not teleportation has been successful (5$^{th}$ column).
If successful, the state onto which atom 1 is projected
($\ket{\alpha_1}$ or $\ket{\alpha_1'}$) is presented (6$^{th}$
column). If unsuccessful, the associated unconditional failure
probability is given (last column). A schematic diagram of the
teleportation protocol is presented in Fig.~\ref{schemino}.
\begin{widetext}
\begin{center}
\begin{table}
\begin{tabular}{||c|c|c|c|c|c|c||}
\colrule \colrule
\bf{Photons} & \bf{Atom 2} & \bf{Path atom 1} & \bf{Path atom 2} & \bf{Teleportation} & \bf{Internal state atom 1} & \bf{Failure probability} \\
\colrule \colrule
2 & -- & -- & -- & Unsuccessful & -- &  $\frac{1}{8}(1+ \cos \vartheta)$\\
\colrule \colrule
$\,$ & $\ket{e_2}$ & -- & -- & Unsuccessful & -- & $\frac{1}{8}(1+ \cos \vartheta)$ \\
\cline{2-7}
$\,$ & $\ket{g_2}$ & $l^-_1$ & $l^-_2$ & Successful & $\cos \frac{\vartheta}{2}  \ket{e_1} +e^{i \varphi} \sin \frac{\vartheta}{2}  \ket{g_1} $ & -- \\
\cline{2-7}
1 & $\ket{g_2}$ & $l^-_1$ & $l^+_2$ & Successful \cite{footnote} & $\cos \frac{\vartheta}{2}  \ket{e_1} -e^{i \varphi} \sin \frac{\vartheta}{2}  \ket{g_1} $ & -- \\
\cline{2-7}
$\,$ & $\ket{g_2}$ & $l^+_1$ & $l^+_2$ & Successful & $\cos \frac{\vartheta}{2}  \ket{e_1} +e^{i \varphi} \sin \frac{\vartheta}{2}  \ket{g_1} $ & -- \\
\cline{2-7}
$\,$ & $\ket{g_2}$ & $l^+_1$ & $l^-_2$ & Successful \cite{footnote} & $\cos \frac{\vartheta}{2}  \ket{e_1} -e^{i \varphi} \sin \frac{\vartheta}{2}  \ket{g_1}$ & -- \\
\colrule \colrule
$\,$ & $\ket{g_2}$ & -- & -- & Unsuccessful & -- & $\frac{1}{4}(1- \cos \vartheta)$ \\

\cline{2-7}
$\,$ & $\ket{e_2}$ & $l^-_1$ & $l^-_2$ & Successful & $\cos \frac{\vartheta}{2}  \ket{e_1} +e^{i \varphi} \sin \frac{\vartheta}{2}  \ket{g_1} $ & -- \\
\cline{2-7}
0 & $\ket{e_2}$ & $l^-_1$ & $l^+_2$ & Successful \cite{footnote} & $\cos \frac{\vartheta}{2}  \ket{e_1} -e^{i \varphi} \sin \frac{\vartheta}{2}  \ket{g_1}$ & -- \\
\cline{2-7}
$\,$ & $\ket{e_2}$ & $l^+_1$ & $l^+_2$ & Successful & $\cos \frac{\vartheta}{2}  \ket{e_1} +e^{i \varphi} \sin \frac{\vartheta}{2}  \ket{g_1} $ & -- \\
\cline{2-7}
$\,$ & $\ket{e_2}$ & $l^+_1$ & $l^-_2$ & Successful \cite{footnote} & $\cos \frac{\vartheta}{2}  \ket{e_1} -e^{i \varphi} \sin \frac{\vartheta}{2}  \ket{g_1}$ & -- \\
\colrule \colrule
\end{tabular}
\caption{\label{table} Teleportation measurement scheme. Each case
is represented by given outcomes of the number of photons (1$^{st}$
column), the internal state of atom 2 (2$^{nd}$ column), and the
paths along which the two atoms are found (3$^{th}$ and 4$^{th}$
columns). In the 5$^{th}$ column it is indicated whether or not
teleportation has been successful. If successful, the state onto
which atom 1 is projected ($\ket{\alpha_1}$ or $\ket{\alpha_1'}$) is
presented (6$^{th}$ column). If unsuccessful, the associated
unconditional failure probability is given in the last column.}
\end{table}
\end{center}
\end{widetext}
The total failure probability, obtained as the sum of the
unconditioned failure probabilities (last column of Table I), is
1/2. Teleportation is thus successful with probability 1/2.

Remarkably, notice that only \emph{local} measurements on
the two atoms and the cavity field are required in order to complete
the teleportation. Direct projections onto highly entangled
states are therefore avoided in our scheme. In Appendix \ref{appB},
we develop a more detailed analysis of the mechanism behind the
scheme.

Finally, unlike previous cavity QED protocols not requiring
direct Bell-state measurements \cite{zheng,ye-guo}, the interaction
time of each atom with the cavity does not need to fulfill any
holonomous constraint. It is only required that it is large enough
in order for (\ref{cond-stati-phi}) to hold with reasonable
approximation.
It should be noted that a problem might arise for the
implementation of the present scheme given that cavity-photon-number
measurements typically require fine tuning of the interaction times
between the field and probe atoms. In Appendix \ref{appC}, we show
how the atomic which-path information can be exploited in order to
accomplish photon-number measurements that require non holonomous
constraints.

\section{Orthogonality of the outgoing atomic wavepackets and which-path information} \label{orthogonality}
In this section, we discuss in more details the features of the
translational states introduced in Eq.~(\ref{stati_phi}) and the
conditions for \emph{which path} information to be accessible.

In the nodal case, using Eqs. (\ref{H-nodal}), (\ref{Ualpha1}),
(\ref{Ualpha2}) and (\ref{stati_phi}), the outgoing translational
wavepackets $\ket{\Phi_{n,i}^{\pm}}$ take the form
\begin{eqnarray}\label{U-N1}
\ket{\Phi_{n,1}^{\pm}}&=&U_{N,1}^{(n,\pm)}\ket{\varphi_1(0)}=\nonumber \\
&=& \exp[i \hbar \frac{\varepsilon^{2} k^{2}}{12 m}(n+1)t_{1}^{3}] \nonumber \\
& &\cdot \exp[\mp
i \varepsilon k\sqrt{n+1}t_{1}(\hat{x}_{1}+\frac{\hat{p}_{1}}{2
m}t_{1})]\ket{\varphi_1(0)},\nonumber \\
\,
\end{eqnarray}
and
\begin{eqnarray}\label{U-N2}
\ket{\Phi_{n,2}^{\pm}}&=&U_{N,2}^{(n,\pm)}\ket{\varphi_2(0)}= \nonumber \\
&=& \exp\{\mp i \varepsilon k \sqrt{n+1}
(t_{3}-t_{2})[\hat{x}_{2}+ \frac{\hat{p}_{2}}{2
m}(t_{3}+t_{2})]\} \nonumber \\
& &\cdot\exp [i \hbar \frac{\varepsilon^{2} k^{2}}{12m}
(n+1)(t_{3}-t_{2})^{3}]\ket{\varphi_2(0)}.
\end{eqnarray}
Using Eqs. (\ref{U-N1}) and (\ref{U-N2}), it can be shown that
\cite{Vag-Cus,epl-2atoms,which-path}
\begin{eqnarray}\label{scalar-prod-N}
\bra{\Phi_{n,i}^{+}}\Phi_{n,i}^{-}\rangle(\tau_i)=\exp \left[-i\left(2\varepsilon k \sqrt{n+1}\,x_{0,i}\right) \tau_i\right] \cdot \nonumber \\
\cdot \exp \left[ -(n+1)\left(\frac{\hbar\varepsilon
k}{m}\right)\left(\frac{\tau_i^2}{8\sigma_{x_{i}}^2}+\frac{4m^2}{8\sigma_{p_{i}}^2}\right)\tau_i^2\right],
\end{eqnarray}
where $x_{0,i}$ stands for the initial average value of the atomic
position along the cavity axis. Eq.~(\ref{scalar-prod-N}) clearly
shows the presence of a damping factor which causes the scalar
products $\bra{\Phi_{n,i}^{+}}\Phi_{n,i}^{-}\rangle$ to vanish at
long times. This proves Eq.~(\ref{cond-stati-phi}) in the nodal
case.

Such behavior, which is at the origin of the non-dissipative damping
of the Rabi oscillations \cite{Vag-Cus,which-path}, arises from the
increasing distance in the phase space \cite{Chian} of the deflected
outgoing components $\ket{\Phi_{n,i}^{\pm}}$ of the incoming
wavepacket $\ket{\varphi_i(0)}$ \cite{Aha}. To better highlight this
phenomenon, Eq.~(\ref{scalar-prod-N}) can indeed be rewritten in the
form \cite{Vag-Cus} (from now on, the subscript $i$ will be omitted
for simplicity)
\begin{eqnarray}\label{scalar-prod-N-2}
\bra{\Phi_{n}^{+}}\Phi_{n}^{-}\rangle(\tau)&=&\exp\left[-i\Omega_{n}(\tau)\tau\right]\nonumber
\exp\left\{-\frac{\left[x_n^+(\tau)-x_n^-(\tau)\right]^2}{8\sigma_x^2}\right.
\nonumber
\\  & &
\left.
-\frac{\left[p_n^+(\tau)-p_n^-(\tau)\right]^2}{8\sigma_p^2}\right\}
\end{eqnarray}
with
\begin{eqnarray}\label{omegan-pn}
\Omega_{n}(\tau)&=&2k\varepsilon \sqrt{n+1}
\left(x_0+\frac{p_0}{2m}\tau\right),\\
x_n^{\pm}(\tau)&=&x_0+\frac{p_0}{m}\tau \mp \frac{\hbar k \epsilon}{2m}\sqrt{n+1}\,\tau^2, \\
p_n^{\pm}(\tau)&=&p_0\mp \hbar k \varepsilon \sqrt{n+1} \,\tau.
\end{eqnarray}
Here $p_0$ stands for the initial average momentum. The above
equations show that wavepackets $\ket{\Phi_{n}^{+}}$ and
$\ket{\Phi_{n}^{-}}$ respectively represent negatively and
positively deflected components of the input wavepacket, the
deflection getting larger as $n$ and/or the atom-cavity interaction
time $\tau$ grow. This is the reason why, when the interaction time
of each atom with the cavity is large enough, \emph{which-path}
information becomes accessible so that the quantum paths associated
with states $\ket{\Phi_{n}^{\pm}}$ can be distinguished (see Sec.
\ref{QT-section}). In order to better illustrate such effect, we
consider an atom of mass $m=10^{-26}$ kg entering a microwave cavity
in a nodal region. Assume that the initial translational state of
the atom is a Gaussian wavepacket of width $\sigma_x=\lambda/10$
($\lambda=2\pi/k=10^{-5}$m) with $x_0=p_0=0$ and that the atom-field
coupling constant $\varepsilon = 10^5$ sec$^{-1}$.  The resulting
quantum paths $l^{\pm}$ associated with wavepackets
$\ket{\Phi_{0}^{\pm}}$ (i.e. those involved in the teleportation
scheme) are shown in Fig.~2 together with their widths
$\sigma_{l^{\pm}}$ (i.e. the standard deviations of
$\abs{\braket{x}{\Phi_{0}^{\pm}}}^2$) as functions of the rescaled
atom-cavity interaction time $\varepsilon \tau$.
\begin{figure}[htbp] \label{Qpath} {\hbox{
{\includegraphics[scale=0.3]{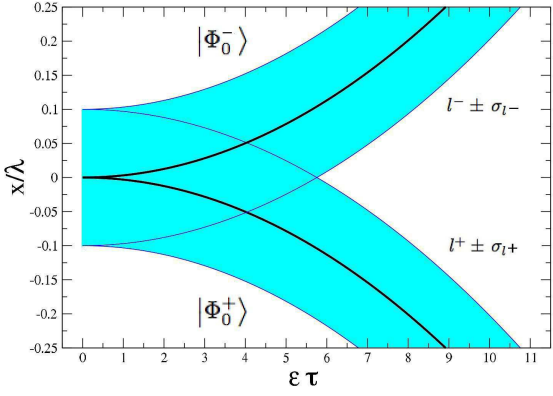}}}} \caption{(Color online)
Quantum paths $l^{+}\pm \sigma_{l^{+}}$ and $l^{-}\pm
\sigma_{l^{-}}$, associated with wavepackets $\ket{\Phi_{0}^{\pm}}$,
versus the rescaled atom-cavity interaction time $\varepsilon \tau$.
The parameters used are: $\lambda=10^{-5}$m, $\varepsilon = 10^5$
sec$^{-1}$, $m=10^{-26}$ kg, $\sigma_x=\lambda/10$ and $x_0=p_0=0$.}
\end{figure}
Notice that the deflection of the two outgoing paths
increase as $\varepsilon \tau$ is raised up to the point that for
atom-cavity interaction times larger than $\simeq 6/\varepsilon$ the
two paths can be reliably distinguished through position
measurements. Even fewer Rabi oscillations are needed in order for
the orthogonality of $\ket{\Phi_{0}^{+}}$ and $\ket{\Phi_{0}^{-}}$
to be achieved. This is shown in Fig.~\ref{D} where the
distinguishability $D$, according to the Englert's definition
\cite{Engl}, is plotted as a function of $\varepsilon \tau$. In the
present case, $D$ take the form \cite{which-path}
\begin{equation}\label{def-D}
D=\sqrt{\left(1-\abs{\braket{\Phi_{0}^{+}}{\Phi_{0}^{-}}}^{2}\right)}.
\end{equation}
\begin{figure}[htbp]  {\hbox{
{\includegraphics[scale=0.3]{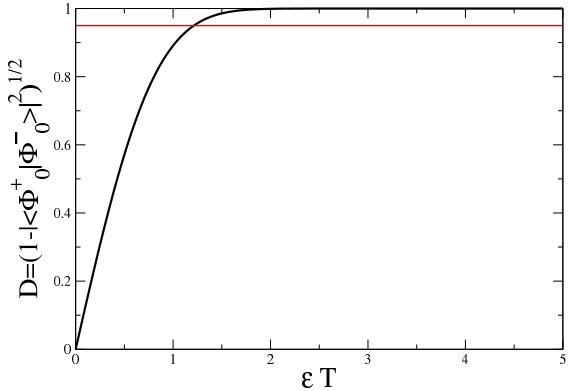}}}} \caption{\label{D}(Color online)
Distinguishability $D$ as a function of the rescaled atom-cavity
interaction time $\varepsilon \tau$. When $D=1$ the which-path information is completely accessible. The parameters used are:
$\lambda=10^{-5}$m, $\varepsilon = 10^5$ sec$^{-1}$, $m=10^{-26}$
kg, $\sigma_x=\lambda/10$ and $x_0=p_0=0$.}
\end{figure}
Notice that $D>95\%$ already for $\varepsilon \tau>1.2$.
The scalar product $\braket{\Phi_{0}^{+}}{\Phi_{0}^{-}}$ therefore
takes less time to vanish ($\simeq 1.2/\varepsilon$) than that
required for distinguishing the atomic position associated with each
path ($\simeq 6/\varepsilon$). The reason of such behaviour is that,
according to Eq.~(\ref{scalar-prod-N-2}), the damping of
$\braket{\Phi_{0}^{+}}{\Phi_{0}^{-}}$ is due to the trajectories in
both the position and momentum space. This suggests that momentum, rather than position, measurements might be more suitable in order to acquire the which-path information for some values of the parameters. Property
(\ref{cond-stati-phi}) holds in the antinodal case as well. Indeed,
using Eqs. (\ref{H-antinodal}), (\ref{Ualpha1}), (\ref{Ualpha2}) and
(\ref{stati_phi}), it turns out that, analogously to the nodal case,
each scalar product $\bra{\Phi_{n}^{+}}\Phi_{n}^{-}\rangle (\tau)$
is always proportional to a damping factor. For instance, in the
case $n=1$ it can be calculated as \cite{epjd}
\begin{eqnarray}
\braket{\Phi_{1}^{+}(\tau)}{\Phi_{1}^{-}(\tau)}=e^{i\frac{\omega_{0}}{2}\tau}e^{-i(a_{1}^2+b_1^2)\frac{\sin(\omega_{0}\tau)}{\cosh(\omega_{0}\tau)}}\cdot \qquad \qquad \qquad \nonumber \\
\cdot e^{\frac{i}{2}\tanh(\omega_{0}\tau)[(a_{1}^2-b_{1}^2)(1+\cos(2\omega_{0}\tau))+2a_{1}b_{1}\sin(2\omega_{0}\tau)]}\cdot\nonumber\\
\cdot\frac{1}{\sqrt{\cosh(\omega_{0}\tau)}}\,e^{-(a_{1}^2+b_1^2)(1-\frac{\cos(\omega_{0}\tau)}
{\cosh(\omega_{0}\tau)})}\cdot \nonumber \\
\cdot e^{-\tanh(\omega_{0}\tau)[a_{1}b_{1}(1-\cos(2\omega_{0}\tau))+\frac{1}{2}(a_{1}^2-b_{1}^2)\sin(2\omega_{0}\tau)]}\nonumber \\
\propto{[1-\frac{(\omega_{0}\tau)^2}{2}]}\cdot\exp{\{-2a_{1}^2(\omega_{0}\tau)^2\}} \,\,\,\,\,\,
(\omega_{0}\tau<1)\quad \nonumber
\end{eqnarray}
where $\omega_{0}^2 =(\hbar k^{2}/m \varepsilon$),
$a_{1}=x_{0}\sqrt{m\omega_{0}/2\hbar}$ and
$b_{1}=(p_{0}/\sqrt{2m\hbar\omega_{0}})$. As in the nodal case, the
damping factor is due to the increasing distance in the phase space
of the deflected components of the incoming wavepacket \cite{epjd}.

\section{Conclusions}\label{conclusions}

In this paper we have presented a scheme for conditionally
teleporting an unknown quantum state between two atoms interacting
in succession with the same cavity mode within the optical
Stern-Gerlach model. Such model, to be regarded as a
generalization of the familiar Jaynes-Cummings Hamiltonian, allows
to account for the atomic translational dynamics. The inclusion of
such dynamics yields the well-known splitting of the wavefunction of
a flying atom into a set of deflected wavepackets. Such phenomenon
could be expected to have a negative effect on quantum information
processing tasks. Indeed, it is known to spoil the non-local
correlations between two atoms which successively interact with the
same cavity mode \cite{epl-2atoms,epjd}. Nonetheless, in this work
we have shown how exactly the above-mentioned splitting can be
fruitfully exploited in order for the atomic translational degrees
of freedom to behave as channels allowing efficient transmission of
quantum information.

Both in the nodal and antinodal case, we have shown that successful
teleportation can be obtained with probability 1/2 by measuring the
number of cavity photons, the internal state of atom 2 and the
position of the two atoms once they are out of the cavity. The
teleportation protocol can be therefore implemented through local
operations. No direct Bell-state measurements are thus necessary in our scheme.

The essential requirement for our protocol to work is that the time
of flight of each atom inside the cavity is sufficiently long in
order \emph{which-path} information to become accessible. Indeed,
the initial wavepacket of each atom splits into a set of outgoing
deflected wavepackets which turn out to be orthogonal, and thus
distinguishable, provided the atom-cavity interaction time is large
enough. Significantly, unlike previous proposals in cavity
QED that do not require direct Bell-state measurements, this
implies a \emph{non holonomous} constraint on the atom-cavity
interaction times. No precise tuning of the atomic flight times
inside the cavity is thus needed.

Nonetheless, it should be observed that, in addition, the
atom-cavity interaction times must be short enough in order for the
lowest-order approximation of the interaction Hamiltonian [Eqs.
(\ref{Hif-nodal}) and (\ref{Hif-antinodal})] to hold for the whole
time of flight of each atom in the cavity. However, this is not a
strong constraint. Interaction times of the order of a few Rabi
oscillations are indeed enough for a \emph{which-path} information
to be accessed (see the numerical example of Fig.~2 where 6 Rabi
oscillations are enough).

To prevent decoherence effects due to the cavity mode
damping, it is of course required the total time of the process
$t_3$ to be shorter than the cavity coherence time $\tau_c$. The
time $t_3$ can be written as $t_3=\tau_1+(t_2-t_1)+\tau_2$, where
$\tau_i$ is the atom-field interaction time for the $i$th atom and
$(t_2-t_1)$ represents the time between the exit of atom 1 and the
entering of atom 2. Since our protocol does not depend on
$(t_2-t_1)$ such time can be made as small as allowed by the
experimental capabilities. It follows that for all practical
purposes it is enough to require that $\tau_1+\tau_2 \ll \tau_c$. As
pointed out above, each $\tau_i$ is required to be larger than a few
Rabi oscillations. This also yields a lower bound for $\tau_c$ that can be however achieved with present-day technology
(see e.g. \cite{raimond,Rempe1992,Hood2001,Mabuchi2002,Vahala2003,Aoki2006,raimond2007}).

Furthermore, notice that, even though the first and the second atom
can be found into, respectively, two and five quantum paths, it is
enough to measure only two paths for each atom ($l_i^{\pm}$
associated with $\ket{\Phi_{0,i}^{\pm}}$) in order to teleport the
initial state of atom 2 into atom 1. As emphasized in Sec.
\ref{QT-section}, the labeling of such two paths is irrelevant given
that it is enough to know only whether the atoms are found in the
same path or not. In the latter case, the teleportation can be
finalized after a 180 degree rotation around the $z$-axis.

Regarding the position measurements of each atom, these should be
performed in such a way not to affect its internal state in the
computational space $\left\{\ket{g},\ket{e}\right\}$. This could be
accomplished by sending light on the atom of wavelength suitable to
excite an atomic transition different from
$\ket{g}\leftrightarrow\ket{e}$.

Finally, this work opens the possibility of exploiting the atomic
translational degrees of freedom in cavity QED in order to perform other
typical quantum information processing tasks, such as the generation
of maximally entangled states.

\begin{acknowledgments}
G. Massimo Palma is gratefully acknowledged for fruitful discussions
and the critical reading of the manuscript. MT wishes to thank
Andreas Buchleitner for fruitful discussions. FC acknowledges
support from PRIN 2006 ``Quantum noise in mesoscopic systems''.
\end{acknowledgments}

\bigskip
\appendix

\section{Derivation of the final state} \label{AppendixA}

In this Appendix, we describe the procedure for obtaining the state
of the system $\ket{\Psi(t\geq t_3)}$ [Eq.~(\ref{expansion-t3})]
after that both the atoms have exited the cavity. According to
Eq.~(\ref{U-alpha}), such state can be obtained through the
successive application of operators $U_{\alpha,1}$ and
$U_{\alpha,2}$ on $\ket{\Psi(0)}$ [Eq.~(\ref{initial-state})]. We
first rewrite the initial state $\ket{\Psi(0)}$
[Eq.~(\ref{initial-state})] in terms of the dressed states of atom 1
by expressing $\ket{e_1}\ket{0}$ as a linear combination of
$\ket{\chi_{0,1}^{+}}$ and $\ket{\chi_{0,1}^{-}}$. This yields
\begin{equation}\label{initial-state-dressed}
\ket{\Psi(0)}=\left(\frac{\ket{\varphi_1(0)}\ket{\chi_{0,1}^{+}}+\ket{\varphi_1(0)}\ket{\chi_{0,1}^{-}}}{\sqrt{2}}
\right)\,\, \ket{\varphi_2(0)}\ket{\alpha_{2}}.
\end{equation}
We now let $U_{\alpha,1}$ act on the initial state (\ref{initial-state-dressed}) to get $\ket{\Psi(t_1)}$ (i. e. the state of the system after
that atom 1 has exited the cavity). By using (\ref{stati_phi2}), we obtain
%
\begin{eqnarray}\label{state-t2}
\ket{\Psi(t_1)}&=&U_{\alpha,1} \ket{\Psi(0)} = \nonumber \\
&=&\frac{\ket{\Phi_{0,1}^+}\ket{\chi_{0,1}^{+}}+\ket{\Phi_{0,1}^-}\ket{\chi_{0,1}^{-}}}{\sqrt{2}}\,\,
\ket{\varphi_2(0)}\ket{\alpha_{2}}.\nonumber \\
&\,\,&
\end{eqnarray}
%
Since in the time interval between $t_1$ and $t_2$, according to Eqs. (\ref{H-nodal}) and (\ref{H-antinodal}), $H_{\alpha}^{I}(t)=0$ ($\alpha=N,A$),
it turns out that $\ket{\Psi(t_2)}=\ket{\Psi(t_1)}$. Before applying $U_{\alpha,2}$ to $\ket{\Psi(t_2)}$ to get $\ket{\Psi(t_3)}$,
it is convenient to rearrange $\ket{\Psi(t_2)}$ as an expansion in the cavity field Fock states as
\begin{widetext}
\begin{equation}\label{state-t2-fock}
\ket{\Psi(t_2)}=\ket{\Psi(t_1)}
=\left[\frac{\ket{\Phi_{0,1}^+}+\ket{\Phi_{0,1}^-}}{2}\ket{e_1}\ket{0}+\frac{\ket{\Phi_{0,1}^+}-\ket{\Phi_{0,1}^-}}{2}\ket{g_1}\ket{1}\right]\,\,
\ket{\varphi_2(0)}\left[\cos
\frac{\vartheta}{2}\ket{e_2}+e^{i\varphi}\sin\frac{\vartheta}{2}\ket{g_2}\right],
\end{equation}
\end{widetext}
Expanding each state $\ket{g_2}\ket{n}$ and $\ket{e_2}\ket{n}$ in
Eq.~(\ref{state-t2-fock}) in terms of $\ket{g_2}\ket{0}$ and of the
dressed states atom 2 $\ket{\chi_{0,2}^{\pm}}$ and
$\ket{\chi_{1,2}^{\pm}}$ and, once $U_{\alpha,2}$ is applied to
$\ket{\Psi(t_2)}$ with the help of Eq.~(\ref{stati_phi2}), the final
state of Eq.~(\ref{expansion-t3}) is obtained.

\section{Insight into the mechanism behind the scheme} \label{appB}

In the present proposal, quantum information is transferred from
atom 2 to atom 1 by using a three-partite continuous-variable (CV)
ancillary system that consists of the translational degrees of
freedom of both the atoms and the cavity field. Immediately before
atom 2 enters the cavity, the state of the system $\ket{\Psi(t_2)}$,
obtained under application of $U_{\alpha,1}$ onto the initial state
(\ref{initial-state}) [cfr. Eq.~(\ref{state-t2}) in Appendix A], can
be put in the form
\begin{widetext}
\begin{eqnarray}
\ket{\Psi(t_2)}&=&\frac{\ket{\varphi_2(0)}}{2\sqrt{2}}\left(\ket{\chi_{0,2}^+}\ket{\Phi_{0,1}^+}\ket{\alpha_1}-\ket{\chi_{0,2}^+}\ket{\Phi_{0,1}^-}\sigma_z\ket{\alpha_1}
-\ket{\chi_{0,2}^-}\ket{\Phi_{0,1}^+}\sigma_z\ket{\alpha_1} + \ket{\chi_{0,2}^-}\ket{\Phi_{0,1}^-}\ket{\alpha_1}\right.\nonumber \\
& &+
\left.\ket{\xi_{0,2}^+}\ket{\Phi_{0,1}^+}\sigma_x\ket{\alpha_1}-\ket{\xi_{0,2}^-}\ket{\Phi_{0,1}^-}\sigma_x\ket{\alpha_1}
+\ket{\xi_{0,2}^-}\ket{\Phi_{0,1}^+}i\sigma_y\ket{\alpha_1} -
\ket{\xi_{0,2}^+}\ket{\Phi_{0,1}^-}i\sigma_y\ket{\alpha_1}\right),\label{psit2-2}
\end{eqnarray}
\end{widetext}
where
$\ket{\xi_{0,2}^\pm}=(\ket{e_2}\ket{1}\pm\ket{g_2}\ket{0})/\sqrt{2}$
are maximally entangled states between the internal degrees of
freedom of atom 2 and the cavity field. Notice that the first line
of Eq. (\ref{psit2-2}) contains terms proportional to either
$\ket{g_2}\ket{1}$ or $\ket{e_2}\ket{0}$, whereas the states
appearing in the second line involve either $\ket{g_2}\ket{0}$ or
$\ket{e_2}\ket{1}$. First, the structure of state (\ref{psit2-2})
shows that, similarly to what has been pointed out in Ref.
\cite{comment-ye-guo}, successful teleportation can, in principle,
be achieved with probability 1. Indeed, provided the interaction
time $\tau_1$ between the atom 1 and the cavity is large enough that
which-path information becomes accessible, $\ket{\Phi_{0,1}^+}$ and
$\ket{\Phi_{0,1}^-}$ become orthogonal [cfr. Sec. \ref{QT-section},
Eq.~(\ref{prop2})]. Therefore the states
$\ket{\chi_{0,2}^{\eta}}\ket{\Phi_{0,1}^{\eta'}}$,
$\ket{\xi_{0,2}^{\eta}}\ket{\Phi_{0,1}^{\eta'}}$ ($\eta,
\eta'=\pm$), each multiplying $\ket{\alpha_1}$ or
$\sigma_{j}\ket{\alpha_1}$ ($j=x,y,z$), form an orthonormal set.
However, direct projections onto maximally entangled states
$\ket{\chi_{0,2}^{\pm}}$ and $\ket{\xi_{0,2}^{\pm}}$ are expected to
be non trivial. In addition, notice that, even assuming the
feasibility of a direct measurement of these states, this would not
be sufficient to complete the teleportation since states
$\ket{\Phi_{0,1}^{\pm}}$ need to be measured as well. For instance,
the first two terms in the right-hand side of Eq.
(\ref{psit2-2}) show that, without measuring the position of atom 1,
a measurement outcome $\ket{\chi_{0,2}^{+}}$ does not
suffice to conclude whether or not a $\pi$-rotation around the
$z$-axis has to be applied to complete the teleportation: if atom 1
is found in path $l_1^-$ this rotation needs to be performed,
whereas if atom 1 is found in path $l_1^+$ it is not required.
Furthermore, notice that, unlike in the scheme of Ref.
\cite{ye-guo}, each state
$\ket{\chi_{0,2}^{\eta}}\ket{\Phi_{0,1}^{\eta'}}$ and
$\ket{\xi_{0,2}^{\eta}}\ket{\Phi_{0,1}^{\eta'}}$ is not entangled
with respect to the internal variables of the atom to be teleported
(\emph{i.e.} atom 1) and the translational degrees of freedom of
both atoms (\emph{i.e.} part of the ancilla). Finally, we point out
that, at this stage, translational degrees of freedom of
atom 2 do not play any role, yet [notice the common factor
$\ket{\varphi_2(0)}$ in Eq. (\ref{psit2-2})]

The difficulty of projecting onto entangled states is overcome by
applying the second unitary transformation $U_{\alpha,2}$.
Translational degrees of freedom of atom 2 are now involved. Using
Eq. (\ref{stati_phi2}), the application of $U_{\alpha,2}$ in fact
accomplishes the following mapping
\begin{eqnarray}
&& U_{\alpha,2}\ket{\chi_{0,2}^{\eta}}\ket{\varphi_2(0)}\ket{\Phi_{0,1}^{\eta'}}=\ket{\chi_{0,2}^{\eta}}\ket{\Phi_{0,2}^{\eta}}\ket{\Phi_{0,1}^{\eta'}}\\
&& U_{\alpha,2}\ket{\xi_{0,2}^{\eta}}\ket{\varphi_2(0)}\ket{\Phi_{0,1}^{\eta'}}=\left[\frac{\eta}{\sqrt{2}}\ket{g_2}\ket{0}\ket{\varphi_2(0)}\right.\,\,\,\nonumber \\
&&\left.
\,\,\,\,\,\,\,\,\,\,\,\,\,\,\,+\frac{1}{2}\left(\ket{\chi_{1,2}^{+}}\ket{\Phi_{1,2}^{+}}+
\ket{\chi_{1,2}^{-}}\ket{\Phi_{1,2}^{-}}\right)\right]\ket{\Phi_{0,1}^{\eta'}}\label{mapping-app-c-2}
\end{eqnarray}
We see that the second unitary transformation $U_{\alpha,2}$ leaves
the states $\ket{\chi_{0,2}^{\pm}}$ unchanged, but, noticeably,
attaches a different wavepacket $\ket{\Phi_{0,2}^{\pm}}$ of atom 2
to each of them. As
$\ket{\chi_{0,2}^{\eta}}=(\ket{e_2}\ket{0}\pm\ket{g}\ket{1})/\sqrt{2}$
and looking at the first line of Eq. (\ref{psit2-2}), it is clear
that now
distinguishing between $\ket{\chi_{0,2}^{-}}$ and
$\ket{\chi_{0,2}^{-}}$ is no longer required to complete the
teleportation. In order to assess whether or not a rotation
$\sigma_z$ has to be applied, it is sufficient to acquire
information about the positions of the two atoms. If they are found
in paths of equal signs the teleportation is completed already,
whereas in the case of paths with opposite signs a further
application of $\sigma_z$ is needed.

The same phenomenon does not occur for states appearing in
the second line of Eq. (\ref{psit2-2}) due to mapping
(\ref{mapping-app-c-2}). Indeed, the application of $U_{\alpha,2}$
changes both the atom-field maximally entangled states
$\ket{\xi_{0,2}^{+}}$ and $\ket{\xi_{0,2}^{-}}$ yielding terms
proportional to either $(\sigma_x+i\sigma_y)\ket{\alpha_1}=2 \cos
\frac{\vartheta}{2}\ket{g_1}$ or
$(\sigma_x-i\sigma_y)\ket{\alpha_1}=2
e^{i\varphi}\sin\frac{\vartheta}{2}\ket{e_1}$. No teleportation is
therefore achievable in this case.

In summary, as the four terms proportional to states
$\ket{\chi_{0,2}^{+}}$ or $\ket{\chi_{0,2}^{-}}$ in Eq.
(\ref{psit2-2}) are those contributing to successful teleportation,
their common factor $(2\sqrt{2})^{-1}$ yields that the probability
of success of the scheme is 1/2.

\section{Photon-number measurements} \label{appC}

In this Appendix, we present a method to perform the necessary
photon-number measurements required by our teleportation protocol.
This task is achieved through position measurements of a third probe
atom $p$.

To complete the protocol, we need to detect the cavity-field Fock
states $\ket{0}$ and $\ket{1}$ at $t\geq t_3$ (see Sec.~III, Table~1
and Fig.~1). According to the initial state (\ref{initial-state})
and due to the conservation of the free energy $\sum_{i=1,2}S_{z,i}+
a^{\dag}a$, the final state in Eq.~(\ref{expansion-t3}) at $t\geq
t_3$ has the form
\begin{equation}\label{stato-finale-appB}
\ket{\psi(t_3)}=\sum_{n=0,2}\ket{c_n}_{12}\ket{n},
\end{equation}
where $\ket{c_n}_{12}$ are states belonging to the overall Hilbert
space of atoms 1 and 2.

Assume that at $t\geq t_3$ the probe atom $p$ is sent through the
cavity in the ground state $\ket{g_{p}}$ and translational
wavepacket $\ket{\varphi_p(0)}$, and that it interacts with the
field for a time $\tau_p$. As $p$ exits the cavity, the final state
of the total system has the form
\begin{equation}\label{stato-finale-cavity-probe}
\ket{\Psi(t_3+\tau_p)}=\sum_{n=0,2}\ket{c_n}_{12}\left[U_{\alpha,p}\ket{\varphi_{p}(0)}\ket{g_p}\ket{n}\right],
\end{equation}
where the evolution operator associated with the atom $p$-field
dynamics $U_{\alpha,p}$ has a form analogous to Eq.~(\ref{Ualpha1})
(the integration bounds in this case are obviously $t_{3}$ and
$\tau_{p}$). By using Eq.~(\ref{stati_phi2}), each state
$\ket{\varphi_{p}(0)}\ket{g_p}\ket{n}$ in the right-hand side of
Eq.~(\ref{stato-finale-cavity-probe}), once expressed in terms of
the dressed states
$\ket{\chi_{n,p}^{\pm}}=\left(\ket{e_{p}}\ket{n}\pm
\ket{g_{p}}\ket{n+1}\right)/\sqrt{2}$, transforms according to
\begin{eqnarray}\label{stati-trasformati-appB}
&U_{\alpha,p}\ket{\varphi_{p}(0)}\ket{g_p}\ket{0}=\ket{\varphi_{p}(0)}\ket{g_p}\ket{0},\label{map0}\\
&U_{\alpha,p}\ket{\varphi_{p}(0)}\ket{g_p}\ket{1}=\frac{\ket{\Phi^+_0}\ket{\chi^+_{0,p}}-\ket{\Phi^-_0}\ket{\chi^-_{0,p}}}{\sqrt{2}},\label{map1}\\
&U_{\alpha,p}\ket{\varphi_{p}(0)}\ket{g_p}\ket{2}=\frac{\ket{\Phi^+_1}\ket{\chi^+_{1,p}}-\ket{\Phi^-_1}\ket{\chi^-_{1,p}}}{\sqrt{2}}.\label{map2}
\end{eqnarray}
Therefore, the unitary operator $U_{\alpha,p}$ in fact maps the Fock
states $\ket{0}$, $\ket{1}$ and $\ket{2}$ into the orthogonal
evolved states (\ref{map0}), (\ref{map1}) and (\ref{map2}),
respectively. A position measurement of the probe atom, in general,
cannot distinguish such states since the wavefunctions
$\ket{\varphi(0)}$,$\ket{\Phi^{\pm}_0(x)}$ and
$\ket{\Phi^{\pm}_1(x)}$ do not form an orthogonal set. However,
provided the interaction time $\varepsilon\tau_p$ is large enough,
these translational states have a negligible overlap according to
\begin{eqnarray}
&\lim_{\tau_p\rightarrow
\infty}\bra{\varphi(0)}\Phi_{n}^{\eta}\rangle =0, \label{cond-stati-phi-2-a}\\
& \lim_{\tau_p\rightarrow
\infty}\bra{\Phi_{n}^{\eta}}\Phi_{n'}^{\eta'}\rangle =0,
\label{cond-stati-phi-2-b}
\end{eqnarray}
with $n=0,1,2,...$ and $\eta,\eta'=\pm$. Notice that
property (\ref{cond-stati-phi}) is a particular case of
Eq.~(\ref{cond-stati-phi-2-b}).

When the probe atom interacts with the cavity field in a
nodal region properties (\ref{cond-stati-phi-2-a}) and
(\ref{cond-stati-phi-2-b}) are explainable as due to the fact that
each translational wavepacket $\ket{\Phi_{n}^{\eta}}$ has an
associated acceleration along the $x$-cavity axis $a_n^{\eta}$ that
depends on both $n$ and $\eta$ according to $a_n^{\eta}=-a_0\,\eta
\sqrt{n+1}$ [$a_0=(\hbar k \epsilon/m)$]\cite{Vag-Cus}. Therefore,
provided $\tau_{p}$ is large enough, 
the wavepackets
$\ket{\Phi_{n}^{\eta}}$ become mutually distinguishable \cite{Vag-Cus}.


\end{document}